\documentclass[aps, twocolumn, superscriptaddress]{revtex4-1}%

\usepackage{amsfonts}
\usepackage{amsmath}
\usepackage{amssymb}
\usepackage{graphicx}
\usepackage{color}%
\usepackage{hyperref}
\usepackage{bm}
\providecommand{\U}[1]{\protect\rule{.1in}{.1in}}

\begin{document}

\title{Discrete time crystal in a finite chain of Rydberg atoms without disorder}

\author{Chu-hui Fan}
\altaffiliation{Contributed equally to this work}
\affiliation{School of Physics, Northeast Normal University, Changchun 130024, China}

\author{D. Rossini}
\altaffiliation{Contributed equally to this work}
\affiliation{Dipartimento di Fisica, Universit\`a di Pisa and INFN, Largo Pontecorvo 3,
I-56127 Pisa, Italy}

\author{Han-Xiao Zhang}
\affiliation{School of Physics, Northeast Normal University, Changchun 130024, China}

\author{Jin-Hui Wu}
\email{jhwu@nenu.edu.cn}
\affiliation{School of Physics, Northeast Normal University, Changchun 130024, China}

\author{M. Artoni}
\affiliation{Department of Engineering and Information Technology and
  Istituto Nazionale di Ottica (INO-CNR), Brescia University, 25133 Brescia, Italy}

\author{G. C. La Rocca}
\email{giuseppe.larocca@sns.it}
\affiliation{Scuola Normale Superiore and CNISM, 56126 Pisa, Italy}

\date{\today}

\begin{abstract}
  We study the collective dynamics of a clean Floquet system of cold atoms, numerically simulating
  two realistic set-ups based on a regular chain of interacting Rydberg atoms driven by laser fields.
  In both cases, the population evolution and its Fourier spectrum display clear signatures of a discrete
  time crystal (DTC), exhibiting the appearance of a robust subharmonic oscillation which persists on a time scale
  increasing with the chain size, within a certain range of control parameters.
  We also characterize how the DTC stability is affected by dissipative processes, typically present
  in this atomic system even though the Rydberg state is very long lived.
\end{abstract}

\maketitle

\section{Introduction}

Spontaneous symmetry breaking is essential to modern physics~\cite{strocchi}. Inspired by the concept
of crystal order in space, in 2012 Frank Wilczek first proposed the idea of a time-crystal phase,
corresponding to spontaneous time-translation-symmetry breaking~\cite{wilczek}, whereby time-periodic properties,
  {\it i.e.}~a clock, emerge in a time-invariant dynamical system.
  This intriguing idea was found to be unfeasible~\cite{bruno,oshikawa} at thermal equilibrium,
  though it can be suitably generalized~\cite{strocchi2}. It was soon realized, however, that periodically driven (a.k.a.~Floquet)
  systems may enter a discrete time-crystal (DTC)
phase~\cite{sacha,sondhi,else,Yao,Khemani}, in which the dynamics is governed by a periodicity
that is different (typically a subharmonic) from that of the Hamiltonian. Signatures of the DTC phase have been
observed in a variety of experimental platforms, e.g., in trapped ions~\cite{zhang}, diamond nitrogen
vacancy centers~\cite{Choi}, superfluid systems~\cite{Autti,smits}, and spin NMR systems~\cite{Pal,Rovny}.
Meanwhile, many extensions of the DTC paradigm have been theoretically discussed as, e.g.,
the possibility to stabilize critical time crystals~\cite{Ho}, prethermal time crystals~\cite{Else2,zeng}, boundary time crystals~\cite{Iemini},
Dicke time crystals~\cite{gong,zhu}, fractional time crystals~\cite{matus}, or even topological time crystals~\cite{lustig,Giergiel}.

The onset of the nonequilibrium DTC phase implies the self-reorganization of a Floquet many-body system, such that the dynamic behavior
of an observable switches to an oscillatory motion characterized by a spontaneously chosen period which differs from that of the Hamiltonian~\cite{rev1,rev2}.
Besides the periodic driving and the many-body interactions, also the presence of disorder in the system was initially assumed to be a prerequisite for the stabilization of the DTC phase, as a consequence of many-body localization~\cite{basko,mbl}.
However, in view of experimental~\cite{Choi, Rovny} and theoretical~\cite{Huang, Russomanno} findings, the occurrence
of a DTC phase has remarkably turned out to be viable even in a clean system, i.e., without disorder, and this case
is recently attracting a great deal of attention~\cite{gambetta,yu,Surace,barfk}. The issue of prethermalization favored by a fast driving,
as well as the detrimental or beneficial role of a dissipative bath, have also been discussed~\cite{Else2,abanin,Lazarides}.

In this context, the investigation of the robustness and limitations of the DTC phase in different systems is clearly
of value. The progress in manipulating cold atoms has made them a unique setting for the quantum simulation of many-body systems~\cite{bloch}, in particular when exploiting Rydberg atom arrays with controllable spatial configurations~\cite{labuhn, kim, barredo, bernien}.
Rydberg atoms of high principal numbers~\cite{gallagher} experience strong dipole-dipole
interactions that can be tailored under appropriate level configurations in order to realize a variety of effects,
including dipole blockade~\cite{Saffman}, dipole anti-blockade~\cite{Amthor}, cooperative
nonlinearity~\cite{Yan,Liu}, quantum many body scars~\cite{turner}, and in-phase/anti-phase dynamics~\cite{Fan}.
Consequently, Rydberg atoms have been proven to be a promising platform for
implementing many quantum tasks such as, e.g., the creation of quantum gates~\cite{Isenhower,Protsenko,Keating},
the generation of entanglement states~\cite{Tian,Rao}, or the realization of photonic devices~\cite{Saffman2,Liu2,cat}.
Recently, new features of the collective dynamics of Rydberg atoms beyond equilibrium states
have been studied under Floquet driving in the presence of disorder~\cite{Potirniche}.

A Floquet system is said to be in the DTC phase when the following three conditions are
satisfied~\cite{Huang, Russomanno}: (i) {\em time-translation symmetry breaking}:  an observable displays
an oscillation with frequency different than that of the driving (typically a lower subharmonic);
(ii) {\em rigidity:} the spontaneously generated frequency is robust against parameter variations;
(iii) {\em persistence}: the time over which the oscillation remains in phase increases
with the system size (temporal long-range order).
Although, in principle, the above conditions should hold in the thermodynamic limit, the consideration
of finite systems is definitely relevant to experiments~\cite{zhang}, while helping in gaining insight into the onset of the DTC phase.

The aim of the present work is the exact numerical simulation of two realistic set-ups without disorder,  based on a finite chain of cold Rydberg atoms. Such a platform could be implemented in experiments~\cite{bloch,labuhn, kim, barredo,bernien} and would provide a flexible tuning
of Floquet driving and of many-body interactions, accompanied by long coherence times. In particular, as sketched in Fig.~\ref{fig1}, the Rydberg atoms, arranged along a ring and coupled via nearest-neighbor dipole-dipole interactions, behave as two-level systems driven by laser fields periodically turned "on" and "off". The first set-up we consider requires only one driving responsible for the Rabi cycling between the ground and the Rydberg states, while the second set-up includes an extra dressing field allowing to control the relevant transition frequency via a light shift which is also turned on and off at the Floquet frequency. Both set-ups can be modeled by
spin Hamiltonians~\cite{rev1,rev2} and our strategy is to numerically analyze, without resorting to approximations, the full quantum dynamical behavior
of finite chains of increasing length (up to $L =14$ atoms) in a regime where the relevant energy scales are all comparable.
We look for clear signatures of the DTC phase that should persist for sufficiently long times,
before other mechanisms such as dissipative processes take over.
Our results suggest the onset of a DTC phase, with features and robustness that may depend
on the Rabi frequency and the detuning of the driving laser fields, as well as on the interatomic interaction strength.
The persistence of a DTC in increasingly larger chains is here characterized as a function of the model parameters.
In both of above the set-ups, although differently sensitive to detuning, definite regularities are unveiled
as a function of the model parameters; these can further be understood by resorting to explicit analytical results
for a simplified few-cycles and few-bodies regime. Finally, the effects of dissipative processes,
which are unavoidable and ultimately due to the decay of the long lived Rydberg states, are addressed as well.

The paper is organized as follows.
We first analyze a system that can be implemented experimentally using a single optical control,
describing the model and the basic theoretical approach in Sect.~\ref{sec:model}
and discussing the numerical results in Sect.~\ref{sec:results}.
We then introduce in Sect.~\ref{sec:model2} a distinct model that, although theoretically simpler,
would also require the use of a second optical pump to periodically shift the energy of the relevant atomic transition
via the AC Stark effect. We discuss the stability of the DTC phase in increasingly large systems
for different parameter ranges in Sect.~\ref{sec:stability} and, finally, how it is affected by dissipative processes
in Sect.~\ref{sec:dissipation}.
Conclusions are presented in Sect.~\ref{sec:concl}, while the two appendices are devoted respectively to details concerning the numerical simulations and to the simplified analytical treatment.

\section{Model and stroboscopic evolution}
\label{sec:model}

\begin{figure}[!b]
  \includegraphics[width=8.5cm]{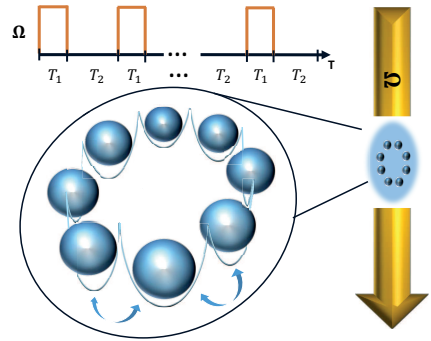}
  \caption{(Color online) Schematic chain of 8 equally spaced Rydberg atoms
    trapped in optical potentials. The atoms interact through a nearest-neighbor coupling $V$,
    and are driven by a common laser field of Rabi frequency $\Omega$.
    Such driving field is periodically switched on ($T_{1}$) and off ($T_{2}$).}
  \label{fig1}
\end{figure}

We consider a closed chain of $L$ equally spaced cold Rydberg atoms tightly trapped in
optical potential wells~\cite{labuhn, kim, barredo}, as illustrated in Fig.~\ref{fig1}.
These atoms can be considered as two-level systems with a common transition frequency $\omega_{gr}$
from their ground $\left\vert g\right\rangle_{j}$ to Rydberg
$\left\vert r\right\rangle_{j}$ states, driven at a Rabi frequency $\Omega$.
They interact through a nearest-neighbor van der Waals (vdW) coupling of strength $V$, while
the driving is alternatively turned on and off in a periodic fashion, so to yield
a binary Floquet Hamiltonian $H(t)$ of period $T=T_{1}+T_{2}$, whose two components are
\begin{widetext}
\begin{align}
  H_{1} = \hbar \sum \limits_{j=1}^{L}
  \left[ \Omega\left( \sigma_{j}^{+} + \sigma_{j}^{-} \right) + \Delta N_{j}^{r} + V N_{j}^{r} N_{j+1}^{r} \right] , &\quad\quad
  {\rm for}\quad (n-1)T \leq t < nT-T_{2}, & \quad {\rm (first \; stage)}
  \label{H1} \\
 {\rm and }\quad\quad H_{2} = \hbar \sum\limits_{j=1}^{L}
  \left( \Delta N_{j}^{r} + V N_{j}^{r}N_{j+1}^{r} \right) , & \quad\quad
  {\rm for}\quad  nT-T_{2}\leq t<nT, & \quad {\rm (second \; stage)}   \label{H2}
\end{align}
\end{widetext}
with $n\in\{1, 2, \ldots, n_{f}\}$ and $n_{f}$ marking the final Floquet cycle.
Here $\sigma_{j}^{+}= |r\rangle_{jj} \langle g|$ ($\sigma_{j}^{-} = |g\rangle_{jj} \langle r|$) refers to the Rydberg
transition raising (lowering) operator, while $N_{j}^{r}=|r\rangle_{jj} \langle r|$
($N_{j}^{g} = |g\rangle_{jj} \langle g|$) depicts the Rydberg- (ground-)
state projection operator for the $j$th atom.
In the above expressions $\Delta=\omega_{d}-\omega_{gr}$ is the detuning of the driving field;
it is worth noting that such detuning also appears in $H_{2}$,
even if the driving is turned off in the second stage, because both $H_{1}$ and $H_{2}$ are written
in the rotating frame with respect to $\omega_{d}$.

The resulting stroboscopic time evolution of the Rydberg chain described above, after $n$
Floquet cycles, is ruled by the unitary operator
\begin{equation}
  U_{F}(n) = \left[ U_{2} U_{1} \right]^{n} = \big[ e^{-iH_{2}T_{2}/\hbar} e^{-iH_{1}T_{1}/\hbar} \big]^{n} ,
  \label{evoop}
\end{equation}
where $U_1 \equiv e^{-iH_{1}T_{1}/\hbar}$ corresponds to the dynamics in the first stage of the
Floquet cycle and $U_2 \equiv e^{-iH_{2}T_{2}/\hbar}$
to that of the second stage.
The Floquet operator $U_F(n)$ depends on the products  \{$\Omega T_{1}$, $\Delta T_{1}$,
$VT_{1}$, $\Delta T_{2}$, $VT_{2}$\}. For the sake of convenience, we
further define $\epsilon$ as a perturbation of Rabi frequency with $\Omega T_{1} = \pi/2 + \epsilon T_{1}$
(the unperturbed value $\Omega T_{1} = \pi/2$ corresponding to a Rabi flip-flop every two Floquet periods)
and hereafter set $T_{1} = 1 \,\mu$s (i.e., $1/T_{1} = 1\,$MHz) thus fixing the unit of time (frequency).

In models describing the onset of DTC signatures~\cite{Yao,Huang}, it is common to turn off the many-body interactions during
the first part of the Floquet cycle, i.e., in $H_{1}$, keeping them only in the second part, i.e., in $H_{2}$.
This procedure seems unfeasible for a realistic Rydberg-atom platform, as it would require to change the relative distance
among atoms rapidly and in a controlled fashion, since the interatomic coupling $V$ depends on the interatomic distance.
A definitely simpler strategy in order to reduce the influence of $V$ in $H_{1}$ is to set $\Omega \gg V$
by applying strong enough driving fields, while setting $T_{1}\ll T_{2}$,
to guarantee that $\Omega T_{1}\approx VT_{2}$, such that the effect of
the internal many-body interactions is comparable to the one of the external driving responsible for the Rabi flipping.

\subsection{Observables}

As typically done in the study of DTC phases, we will present our results in terms of three
  related observables~\cite{Choi,Yao,Ho,Huang}, namely the population difference $P$, its Fourier transform $S$ and a binary order parameter $Q$, that serve to assess the onset of a DTC phase.  The Rydberg chain
 atomic population difference (i.e., the average population imbalance between Rydberg and ground states), is defined as:
\begin{equation}
  P(n) = \frac{1}{L} \sum\limits_{j=1}^{L} \langle \Psi(n)| N_{j} |\Psi(n)\rangle ,
  \label{Pn}
\end{equation}
where $\left\vert \Psi(n)\right\rangle =$ $U_{F}(n)\left\vert \Psi (0)\right\rangle$
denotes the collective state after $n$ Floquet cycles described above, starting from the initial state
$\left\vert \Psi(0)\right\rangle =\left\vert g\right\rangle_1\left\vert g\right\rangle_2...\left\vert g\right\rangle_L$ which is the natural choice for this experimental platform,
while $N_{j}=N_{j}^{r}-N_{j}^{g}$ is the population difference operator for the $j$th atom.

In the case of a large enough number of Floquet cycles, $n_f \gg 1$, the DTC features can be also
assessed by looking at the population difference Fourier spectrum~\cite{Choi,Yao}
\begin{equation}
  S(\nu) = \frac{1}{n_f} \sum\limits_{n=1}^{n_f} P(n) \exp(2\pi i n \nu) ,
  \label{spectra}
\end{equation}
which provides a simple picture on the time-translation-symmetry breaking,
by showing one or more resonant peaks (in the above equation $n_f$ is the total number of Floquet cycles,
and its inverse stands for a normalization coefficient).

Since $P(n)$ oscillates continuously between positive and negative values with two extreme $\pm 1$,
in order to check the stability of the DTC phase, we introduce the binary order parameter~\cite{Ho}
\begin{equation}
  Q(n) = {\rm sgn} \left[ (-1)^{n}P(n)\right] .
  \label{order}
\end{equation}
With our choice of initial state, in the DTC phase of $2T$ periodicity, one finds a constant value $Q(1) = Q(2) = \cdots = Q(n_{f}) = -1$,
indicating that the oscillation frequency is fixed during the whole evolution process.
Otherwise, $Q(n) = \pm 1$ will appear alternatively for identical or distinct numbers of Floquet cycles,
indicating that two or more oscillation frequencies exist before the evolution process stops.

\subsection{Experimental feasibility}

Our model can be realized with present-day cold-atom technology~\cite{labuhn, kim, barredo, bernien}, in a three-level ladder configuration
for $^{87}$Rb atoms with, {\it e.g.}, the Rydberg state $|r\rangle =| 60s, \, j=1/2, \, m_{j}=1/2 \rangle$,
the excited state $|e\rangle = | 5P_{1/2}, \, F=1, \, m_{F}=0 \rangle$,
and the ground state $|g\rangle =| 5S_{1/2}, \, F=1, \, m_{F}=1 \rangle$~\cite{gallagher, morsch,brow}. Experimentally, the Rydberg state $|r\rangle$ is populated from the ground state $|g\rangle$ via a two-photon transition nearly resonant with the intermediate state  $|e\rangle$, but
the latter state can be adiabatically eliminated
in the case of a small two-photon detuning $\Delta$ and two large single-photon detunings, so that the
three-level ladder configuration can be effectively described by our two-level model (referring to states $|g\rangle$ and $|r\rangle$, driven by a single generalized
Rabi frequency $\Omega$ with the detuning $\Delta$)~\cite{morsch,linskens}.

The upper Rydberg state $|r\rangle $ has a decay rate $\Gamma \simeq 2\pi\times 2.0$ kHz
and a vdW coefficient $C_{6} \simeq 2\pi\times 1.4 \times 10^{11}$ $s^{-1}\mu m^{6}$,
yielding the vdW potential $V = C_{6}/R^{6}\in\{0.3 \, {\rm MHz},0.005 \, {\rm MHz} \}$
for a moderate interatomic distance $R \in \{ 12 \, \mu{\rm m}, 24 \, \mu{\rm m} \}$~\cite{gallagher,brow,Saffman}.
Here, we take into account only nearest neighbor atom-atom interactions with $V\approx 0.1$ MHz, and initially disregard
all dissipative effects. However, since the decay rate of the Rydberg state introduces an unavoidable source of dissipation,
its consequences will be considered in Sect.~\ref{sec:dissipation}, in order to assess realistic limits for our simulations.

\section{Results}
\label{sec:results}

Let us now present the outcomes of some numerical exact calculations, aimed at revealing
whether our model of Rydberg chain, described by the stroboscopic time evolution
of the Hamiltonian in Eqs.~\eqref{H1} and~\eqref{H2}, can exhibit the typical DTC features (see also
Appendix~\ref{app:numerics} for details).
Specifically, we shall consider $\epsilon$ and $\Delta$ as two kinds of perturbation
in the presence of a non vanishing $V$, and examine their effects on the
population difference $P(n)$, its Fourier spectrum $S(\nu)$,
and the corresponding order parameter $Q(n)$.
A qualitative insight on some of our findings can also be gained through a direct analysis of the few-cycle dynamics
for two and three atoms (the corresponding analytic expressions of $P(n)$ being provided in Appendix~\ref{app:analytic}).

%%%%%%%%%%%%%%%%%%%%%%%%%%%%%%%%%%%%%%%%%%%%%%%%%%%%%%%%%%%%%%%%%%%%%%%%%%%%%%%%%%%%%%%
\begin{figure}[!t]
  \includegraphics[width=8.5cm]{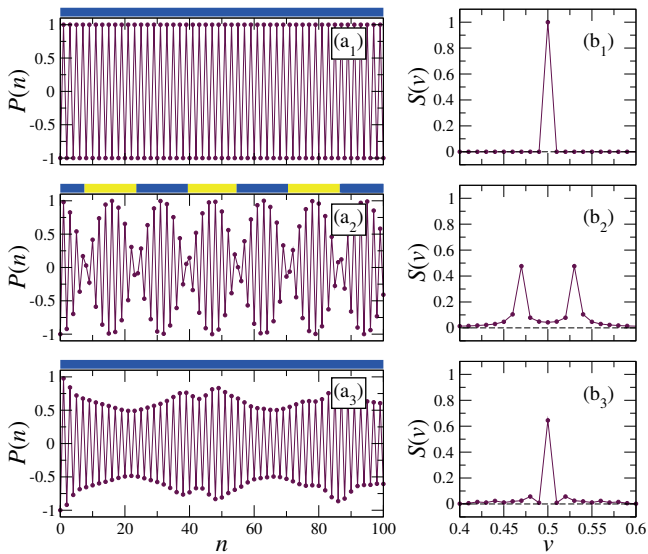}
  \caption{(Color online) The average population difference $P(n)$ (left panels) and corresponding Fourier spectra
    $S(\nu)$ (right panels) for a closed chain of $L=8$ Rydberg atoms, in the case of
    zero detuning ($\Delta = 0$). The various panels stand for $\epsilon = V = 0$ [(a$_{1}$) and (b$_{1}$)];
    $\epsilon = 0.1 \,$MHz and $V=0$ [(a$_{2}$) and (b$_{2}$)];
    $\epsilon = V = 0.1 \,$MHz, and $T_{2}=10 \,\mu$s [(a$_{3}$) and (b$_{3}$)].
    Here and in the next figures, the order parameter $Q(n)$ is plotted in colorbars over the left panels,
    with $Q(n)=-1$ and $Q(n)=1$ shown in blue and yellow, respectively.}
  \label{fig2}
\end{figure}
%%%%%%%%%%%%%%%%%%%%%%%%%%%%%%%%%%%%%%%%%%%%%%%%%%%%%%%%%%%%%%%%%%%%%%%%%%%%%%%%%%%%%%%

We start by taking  $\Delta=0$ in Fig.~\ref{fig2}.
For $\epsilon=0$ and $V=0$ [panels (a$_{1}$) and (b$_{1}$)], the atomic polarization trivially exhibits
perfect Rabi flips between $P(n) = \pm 1$ and $P(n+1) = \mp1 $ in any two consecutive
Floquet cycles, indicating a temporal periodicity which is twice as that of $H(t)$,
and further verified by an invariant order parameter $Q(n)=-1$ and a single subharmonic peak at $\nu=0.5$ in $S(\nu)$.
Since $Q$ is a binary variable, in all the figures
we represent it pictorially with a blue/yellow colorbar over the corresponding panels.
The results become very different in the presence of a nonzero $\epsilon$
[panels (a$_{2}$) and (b$_{2}$) of Fig.~\ref{fig2}]: the perturbation $\epsilon$ indeed results in an excessive
or inadequate rotation, so that $P(n)$ exhibits beating-like oscillations of period
$n_{b} = \pi/( 2|\epsilon|)$
(in the above panels we fixed $\epsilon = 0.1$, and thus $n_b \approx 16$).
As a consequence, $Q(n)$ periodically switches between $-1$ and $1$, while $S(\nu)$ displays
two symmetric subharmonic peaks around $\nu=0.5$.
The population imbalance $P(n)$ no longer exhibits a period twice as that of $H(t)$,
and the perfect temporal periodicity is lost.
Most interestingly, by taking nonzero values both for $\epsilon$ and for $V$,
it is possible to recover results that closely resemble those attained for $\epsilon=V=0$.
Panels (a$_3$) and (b$_3$) of Fig.~\ref{fig2} (where we set  $\epsilon = V = 0.1 \,$MHz and $T_2 = 10 \, \mu$s)
indeed demonstrate this feature, although the specific value reached by $P(n)$,
generally differing from $\pm 1$, now becomes $n$-dependent.
The corresponding Fourier spectrum acquires two small branch peaks around the previous high central one.
This behavior is a signature of a DTC regime induced by the presence of many-body interactions,
the period $2T$ being robust against variations of  $\epsilon$.

%%%%%%%%%%%%%%%%%%%%%%%%%%%%%%%%%%%%%%%%%%%%%%%%%%%%%%%%%%%%%%%%%%%%%%%%%%%%%%%%%%%%%%%
\begin{figure}[ptb]
  \includegraphics[width=8.5cm]{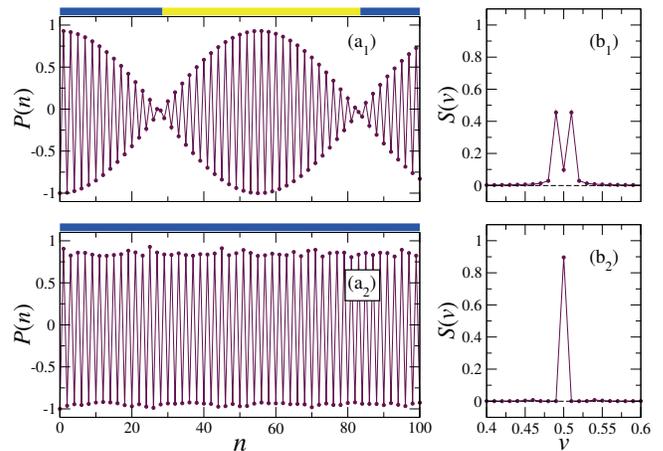}
  \caption{(Color online) Same as in Fig.~\ref{fig2}, but for $\Delta = 0.6 \,$MHz
    and in the case of zero perturbation of the Rabi frequency $\epsilon = 0$ (i.e., $\Omega T_1 = \pi/2$).
    Panels (a$_{1}$) and (b$_{1}$) are for $V=0$, while panels (a$_{2}$) and (b$_{2}$) are
    for $V=0.1 \,$MHz and $T_{2}=10 \, \mu$s.}
  \label{fig3}
\end{figure}
%%%%%%%%%%%%%%%%%%%%%%%%%%%%%%%%%%%%%%%%%%%%%%%%%%%%%%%%%%%%%%%%%%%%%%%%%%%%%%%%%%%%%%%

We proceed by taking a finite detuning $\Delta$ in Fig.~\ref{fig3}.
Notice, however, that in our case $\Delta$ is always kept uniform and does not introduce any disorder in the system.
Panels (a$_{1}$) and (b$_{1}$) of Fig.~\ref{fig3} show that, in the presence of a detuning $\Delta = 0.6 \,$MHz
and switching off the vdW coupling $V$, the atomic polarization exhibits beating-like
oscillations, as expected for independent atoms.
Each beating period can be estimated as the number $n_b$ of Floquet cycles that are needed to recover the same values
for any observable in the perfect system ($\Delta=0$) and in the detuning-perturbed system.
Defining the effective Rabi frequency for the perturbed system $\Omega_{e} \equiv \sqrt{\Omega^{2}+\Delta^{2}}$,
the above condition is thus enforced by requiring $n_b \Omega_e T_1 = n_b \Omega T_1 + 2 \pi$,
which corresponds to $n_{b} \approx 57$ cycles for $\Delta = 0.6 \,$MHz
[see the length of the yellow colorbar in panel (a$_1$)].
Accordingly, $Q(n)$ periodically switches between its extremal values $\pm 1$,
and $S(\nu)$ displays two symmetric peaks around $\nu=0.5$.
Similarly to what observed in panels (a$_3$) and (b$_3$) of Fig.~\ref{fig2}, even in this case we find that
vdW interactions are able to stabilize the DTC regime, the period $2T$ being robust against variations of $\Delta$.
This is explicitly shown in panels (a$_2$) and (b$_2$) of Fig.~\ref{fig3}, for $V=0.1 \,$MHz and $T_2 = 10 \, \mu$s.

%%%%%%%%%%%%%%%%%%%%%%%%%%%%%%%%%%%%%%%%%%%%%%%%%%%%%%%%%%%%%%%%%%%%%%%%%%%%%%%%%%%%%%%
\begin{figure}[ptb]
  \includegraphics[width=8.5cm]{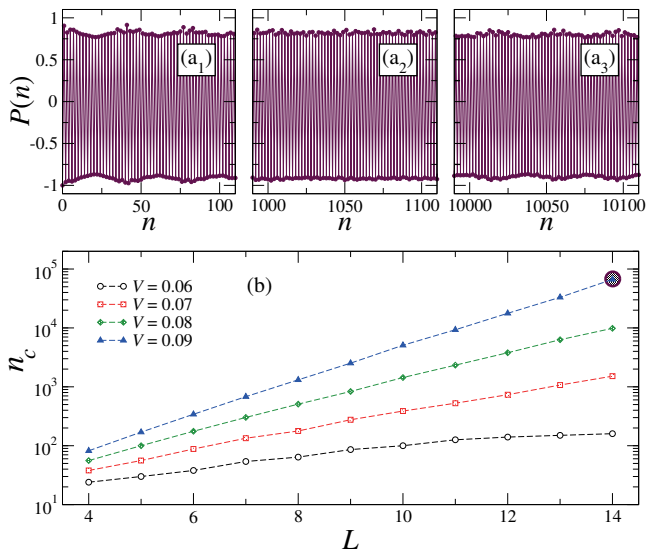}
  \caption{(Color online) (a$_{1}$)-(a$_{3}$): The average population difference
    in various Floquet time intervals, for a closed chain of $L=14$ Rydberg atoms with $V=0.09 \,$MHz.
    (b): The critical oscillation number $n_{c}$ against the size $L$ in semilog scale, for different values of $V$,
    as indicated in the legend. The data point evidenced with a dark circle indicates the situation
    highlighted in the three upper panels.
    The other parameters are chosen as $\Delta=0.6 \,$MHz, $\epsilon=0$, and $T_{2}=15 \, \mu$s.}
  \label{fig4}
\end{figure}
%%%%%%%%%%%%%%%%%%%%%%%%%%%%%%%%%%%%%%%%%%%%%%%%%%%%%%%%%%%%%%%%%%%%%%%%%%%%%%%%%%%%%%%

We now tackle the issue of persistence, which is one of the distinguishing properties of a DTC phase with increasing system size.
Within a certain range of parameters and for a sufficiently large value of $L$,
it is possible to observe a remarkable stability of the oscillations of $P(n)$ over tens of thousands
of Floquet cycles. This is the case for the situation presented in Fig.~\ref{fig4}, panels (a$_{1}$) to (a$_{3}$),
where we simulated a chain with $L=14$ Rydberg atoms.
We may conclude that, in such situation, all the three typical DTC features can be observed in a finite
interacting Rydberg chain, no matter how its many-body Hamiltonian is perturbed through
the Rabi frequency $\epsilon$ or through the detuning $\Delta$.
To be more accurate, a scaling analysis of the persistence of the DTC with the system size $L$ should be performed.
In this respect a useful quantity is the critical Floquet oscillation number $n_{c}$,
which can be defined as the total number of Floquet cycles before $Q(n)$ first changes its sign.
This corresponds to the situation in which
$Q(n)$ first switches from $-1$ to $+1$ (i.e., the smallest value of $n$ for which the colorbar
changes color from blue to yellow).
The panel (b) of Fig.~\ref{fig4} shows that $n_{c}$ increases exponentially with $L$,
at a rate which is sensitive to the strength $V$ of the vdW potential.
It is remarkable that, already with a relatively small number of atoms (of the order of ten),
a sufficiently strong vdW coupling can stabilize a DTC phase over several thousands of Floquet cycles,
thus allowing for a direct experimental test even for a moderate number of Rydberg atoms.

%%%%%%%%%%%%%%%%%%%%%%%%%%%%%%%%%%%%%%%%%%%%%%%%%%%%%%%%%%%%%%%%%%%%%%%%%%%%%%%%%%%%%%%
\begin{figure*}[!t]
  \includegraphics[width=12.5cm]{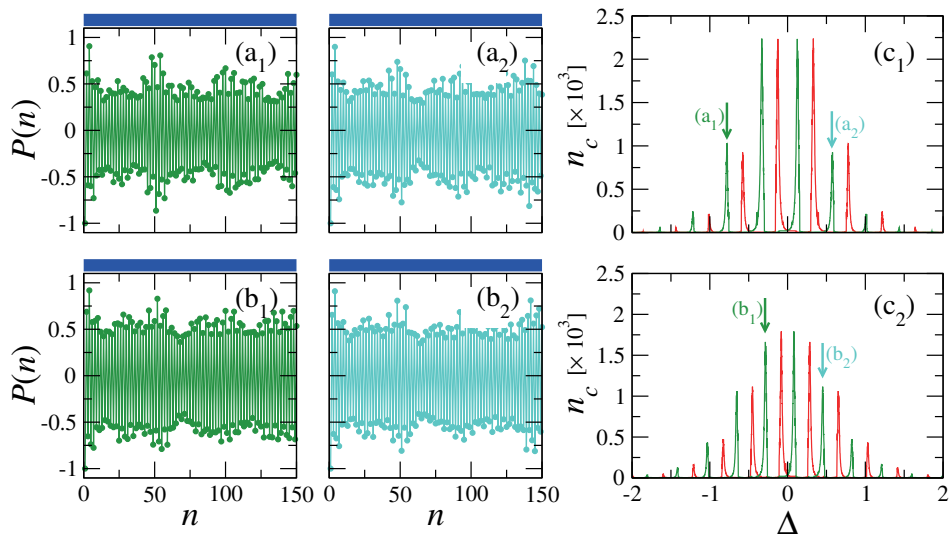}
  \caption{(Color online)
    The average population difference and the critical oscillation number
    for different values of $\epsilon$, $\Delta$, and $V$.
    The three upper panels are for $\epsilon = 0.4 \,$MHz, while the three lower panels are for $\epsilon = -0.4 \,$MHz.
    (a$_{1}$)-(a$_{2}$): The population difference $P(n)$ for $\Delta = -0.782 \,$MHz and $\Delta=0.578 \,$MHz respectively,
    as indicated by arrows in panel (c$_1$).
    (b$_{1}$)-(b$_{2}$): The population difference $P(n)$ for $\Delta = -0.285 \,$MHz and $\Delta = 0.452 \,$MHz respectively,
    as indicated by arrows in panel (c$_2$).
    (c$_1$)-(c$_2$): The critical oscillation number $n_{c}$ as a function of the detuning $\Delta$.
    Green data sets stand for $V = 0.1 \,$MHz, as is the case for all the four panels on the left
    (arrows point to specific values of $\Delta$ on the green data sets).
    The red data sets stand for $V = -0.1 \,$MHz.
    In this figure we considered a chain of $L = 8$ atoms and fixed $T_{2} = 15 \, \mu$s.}
  \label{fig5}
\end{figure*}
%%%%%%%%%%%%%%%%%%%%%%%%%%%%%%%%%%%%%%%%%%%%%%%%%%%%%%%%%%%%%%%%%%%%%%%%%%%%%%%%%%%%%%%

We finally explore the dependence of the DTC phase on both the detuning $\Delta$
and a nonzero Rabi frequency perturbation $\epsilon$.
Panels (c$_{1}$) and (c$_{2}$) of Fig.~\ref{fig5} reveal a number of interesting features that we briefly summarize here.
First of all, the critical oscillation number $n_{c}$ exhibits discrete sharp peaks at specific regularly spaced detunings
with amplitudes decaying relatively fast for increasing $|\Delta|$.
Secondly, we notice that the spacing and the amplitude of these peaks are somewhat different for positive
[panel (c$_1$)] and for negative [panel (c$_2$)] perturbations of the Rabi frequency;
in particular the central peaks are higher for positive $\epsilon$.
While the figure displays data for $|\epsilon| = 0.4 \,$MHz, we have checked that a qualitatively analogous
trend can be found for generic values of $|\epsilon|$.
Panels (a$_{1}$)-(a$_{2}$) and (b$_{1}$)-(b$_{2}$) further show that $P(n)$ maintains stable
oscillations, although with fast amplitude fluctuations that usually get suppressed with increasing $L$ (not shown).
Finally, values of $n_{c}$ attained with a positive $V$ [green data sets in panels (c$_1$) and (c$_2$)]
and those attained with a negative $V$ [red data sets in panels (c$_1$) and (c$_2$)]
display an exact symmetry with respect to $\Delta = 0$,
indicating invariant results for the simultaneous sign change of $V$ and $\Delta$.
Such symmetry holds for any finite values of $V$ and $\Delta$,
and can be understood, as discussed in the Appendix~\ref{app:analytic},
via some analytical insight on the few-cycle dynamics for two and three atoms.

\section{Improved model with an externally controlled detuning}
\label{sec:model2}

In the previous Section we have found that the chain of Rydberg atoms introduced in Sect.~\ref{sec:model}
is able to stabilize a DTC phase, with a critical oscillation number $n_c$ which is rather sensitive to the detuning.
Indeed, as shown in Fig.~\ref{fig5}(c$_1$) and (c$_2$), a small deviation of $\Delta$
from the optimal value may result in a large reduction of $n_{c}$ (as the peaks are quite narrow).
It would be desirable to find a practical way to quench such sensitivity
which may hinder the detection of DTC signatures.

To this purpose, we now propose an alternative model to that defined in Eqs.~\eqref{H1} and~\eqref{H2}.
Specifically one may choose to compensate the detuning $\Delta$ appearing in the rotating frame expression of $H_2$ through the
AC Stark effect (or Autler-Townes effect), by turning on an additional optical pump only
in the second stage of the Floquet cycle, for $nT-T_2\leq t < nT$, which dresses the atom coupling the ground level $|g\rangle$
to an excited level $|e^\prime\rangle$~\cite{cohen}.
This additional pump can be red or blue detuned, the corresponding light shift compensating for $\Delta$
of either positive or negative sign, with a detuning absolute value large enough to allow for the adiabatic elimination
of level $|e^\prime\rangle$. In such case, the Hamiltonian of Eq.~\eqref{H2} simplifies to
\begin{equation}
  \tilde H_{2} = \hbar \sum\limits_{j=1}^{L} V N_{j}^{r} N_{j+1}^{r}.
  \label{H21}
\end{equation}
The new stroboscopic time evolution of this model is thus given by the unitary operator
$\tilde U_{F}(n) = \big[ \exp( -i \tilde H_{2}T_{2}/\hbar) \exp (-iH_{1}T_{1}/\hbar) \big]^{n}$.
We will refer to this as the improved model, as compared to the original model discussed before in Sect.~\ref{sec:model}.
Since the two models are identical in the case of $\Delta=0$, we examine the effect
of a nonzero detuning $\Delta$, now present only in $H_1$, on the critical value $n_c$.

%%%%%%%%%%%%%%%%%%%%%%%%%%%%%%%%%%%%%%%%%%%%%%%%%%%%%%%%%%%%%%%%%%%%%%%%%%%%%%%%%%%%%%%
\begin{figure}[!t]
  \includegraphics[width=8.5cm]{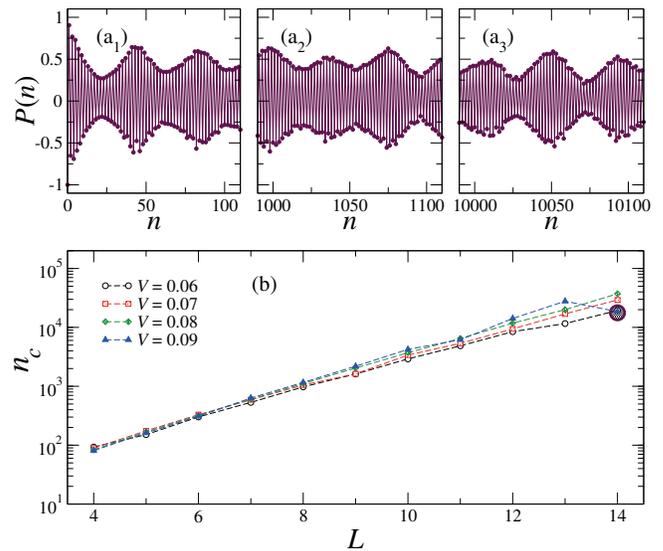}
  \caption{(Color online) Same as in Fig.~\ref{fig4}, but for the alternative model
    that compensates the term in $\Delta$ appearing in Eq.~\eqref{H2} [cf., Eq.~(\ref{H21})].
    All the various parameters are set as in Fig.~\ref{fig4}.
    Notice the reduced sensitivity of $n_c$ to the vdW potential $V$, as compared with the previous model of Sect.~\ref{sec:model}.}
  \label{fig6}
\end{figure}
%%%%%%%%%%%%%%%%%%%%%%%%%%%%%%%%%%%%%%%%%%%%%%%%%%%%%%%%%%%%%%%%%%%%%%%%%%%%%%%%%%%%%%%

%%%%%%%%%%%%%%%%%%%%%%%%%%%%%%%%%%%%%%%%%%%%%%%%%%%%%%%%%%%%%%%%%%%%%%%%%%%%%%%%%%%%%%%
\begin{figure*}[!t]
  \includegraphics[width=16.5cm]{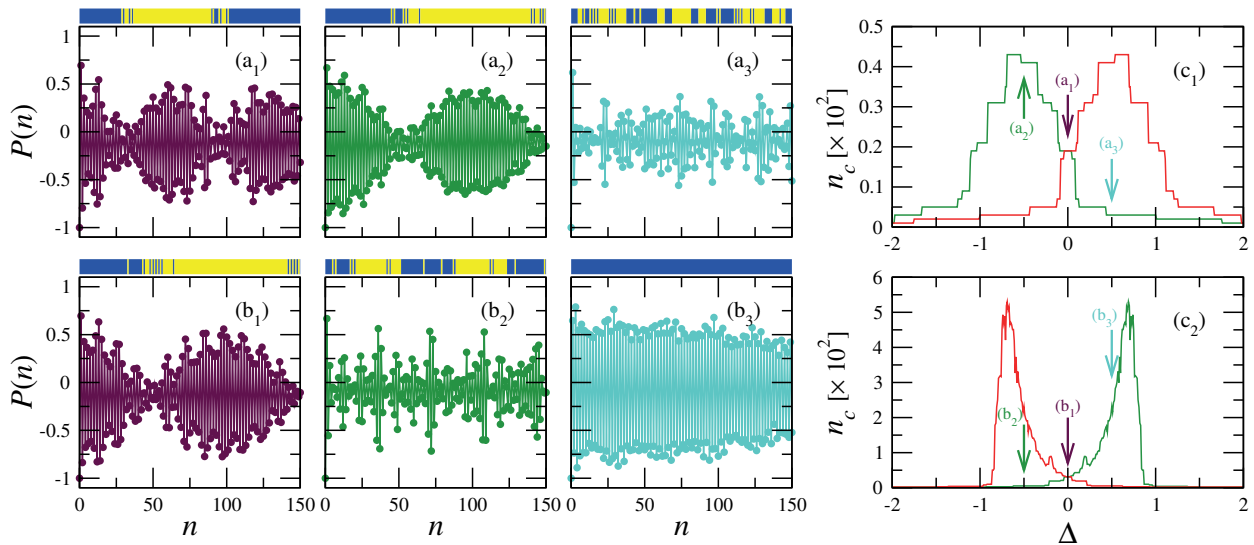}
  \caption{(Color online) Same as in Fig.~\ref{fig5}, but for the improved Rydberg-atom model [cf., Sect~\ref{sec:model2}].
    Upper panels are for $\epsilon = 0.4 \,$MHz, while lower panels are for $\epsilon = -0.4 \,$MHz.
    Panels (a$_{1}$) and (b$_{1}$) are for $\Delta =  0   \,$MHz;
    panels (a$_{2}$) and (b$_{2}$) are for $\Delta = -0.5 \,$MHz;
    panels (a$_{3}$) and (b$_{3}$) are for $\Delta =  0.5 \,$MHz.
    All the plots with $P(n)$ correspond to specific points in the green curves of panels (c$_{1}$)
    and (c$_{2}$) indicated by arrows. Green data sets in panels (c$_{1}$) and (c$_{2}$) stand
    for $V = 0.1 \,$MHz; red data sets stand for $V = -0.1 \,$MHz.
    In this figure we considered a chain of $L = 8$ atoms and fixed $T_{2} = 15 \, \mu$s.}
  \label{fig7}
\end{figure*}
%%%%%%%%%%%%%%%%%%%%%%%%%%%%%%%%%%%%%%%%%%%%%%%%%%%%%%%%%%%%%%%%%%%%%%%%%%%%%%%%%%%%%%%

Figure~\ref{fig6}, plotted in the same way as Fig.~\ref{fig4}, shows that,
also in this alternative model, the DTC features can be stabilized over
several thousands of Floquet cycles.
Indeed, panels (a$_{1}$)-(a$_{3}$) display the time traces of the population imbalance $P(n)$
in three windows with time scales differing by orders of magnitude, for $L=14$ and $\Delta=0.6 \,$MHz. From there, we can see that
fixed-frequency oscillations of $P(n)$ persist over a very long time with smooth amplitude
fluctuations, which are only slightly suppressed at large times. The critical oscillation number $n_{c}$
also increases exponentially as the chain size $L$ grows, and, compared to the original model,
is much less sensitive to the vdW potential $V$ [compare panels (b) of Figs.~\ref{fig4} and~\ref{fig6}].

Let us now discuss more in detail the effects of the detuning $\Delta$ on the rigidity to perturbations
of the DTC phase. To this aim, it is instructive to have a look at panels (c$_{1}$) and (c$_2$) of Fig.~\ref{fig7},
plotted in the same way as the corresponding panels in Fig.~\ref{fig5}.
From there we observe that $n_{c}$ still exhibits an exact symmetry with respect to $\Delta=0$ for two opposite values
of $V$ (as expected again from the analytical results discussed in the Appendix~\ref{app:analytic}).
However, in contrast to the model of Sect.~\ref{sec:model}, the dependence of $n_{c}$ on $\Delta$
yields a smooth envelope rather than distinct sharp peaks.
It is also worth noting that such smooth envelope becomes narrower and higher if the sign
of $\epsilon$ is changed from positive [panel (c$_1$)] to negative [panel (c$_2$)]. This is because the
perturbation due to $\Delta$, here present only in $H_1$, can be partly suppressed (enhanced) by a negative (positive) $\epsilon$
in the expression for $\Omega_{e}=\sqrt{ \big[ \pi/(2T_1)+\epsilon \big]^{2}+\Delta^{2}}$, so that the DTC
phase becomes more (less) stable. However, even for this alternative model, the values of $\Delta$ and $\epsilon$ also affect the role of the many-body interactions as the latter depend on the population of the Rydberg level, and there is a complex interplay among $\Delta$, $\epsilon$ and $V$.
Panels (a$_{1}$)-(a$_{3}$) [resp.~(b$_{1}$)-(b$_{3}$)] further show $P(n)$ for three points
at different positions of the smooth envelope in panel (c$_{1}$) [resp.~(c$_{2}$)], from which we can see
that a larger $n_{c}$ is always accompanied by more stable oscillations.

All in all, the DTC signatures may be more accessible in experiments implementing the
  improved model introduced in this Section, owing to the reduced sensitivity on $\Delta$.
  Clearly, this would require a suitable choice of the additional dressing laser field
  in the second stage of the Floquet cycle.

\section{Parameter space analysis}
\label{sec:stability}

As mentioned in the Introduction, the DTC features should be assessed in the thermodynamic limit $L\rightarrow \infty$,
while our numerical simulations are limited to finite-size chains (yet of experimental relevance).
However, the parameter ranges within which $n_{c}$ increases exponentially with $L$ [as those in Fig.~\ref{fig4}(b)
  and Fig.~\ref{fig6}(b)] may represent a genuine DTC regime.

A different question regards the dependence of the DTC persistence on the parameter range, as another signature of the DTC regime
is its rigidity, i.e., the robustness against parameter variations.
Indeed in Fig.~\ref{fig4}(b) and Fig.~\ref{fig6}(b) we have $\Delta\ne 0$ and fixed $\epsilon=0$;
in such case, the exponential increase of $n_c$ with $L$ is numerically verified
(at least for the sizes we were able to address numerically).
However we have found a quite robust numerical evidence that, for $\epsilon\ne 0$, this conclusion holds true
only within certain ranges of $\Delta$ values.
Incidentally we note that such ranges of parameters are more restrictive for the model of Sect.~\ref{sec:model},
rather than those for the model of Sect.~\ref{sec:model2}.
Below we comment on this rather delicate issue.

\begin{figure}[!t]
  \includegraphics[width=8.5cm]{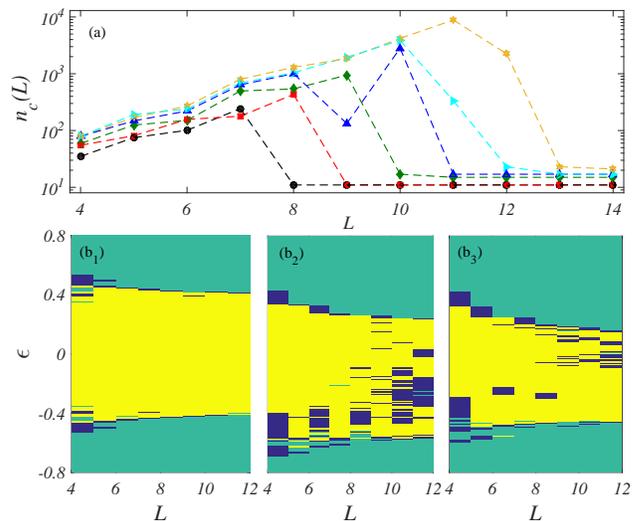}
  \caption{(Color online) (a) The critical oscillation number $n_{c}$ against the chain size $L$,
    for the model of Sect.~\ref{sec:model} with $\Delta=0.16 \,$MHz (black circles), $0.155 \,$MHz (red squares),
    $0.15 \,$MHz (green diamonds), $0.145 \,$MHz (blue triangles up),
    $0.143 \,$MHz (cyan triangles right), and $0.142 \,$MHz (orange stars), respectively. The other parameters
    are chosen as $\epsilon=0.4 \,$MHz, $V=0.1 \,$MHz, and $T_{2}=15 \, \mu$s.
    (b$_1$)-(b$_3$) Finite-size phase diagram of the simplified model in Appendix~\ref{app:analytic}
    [$\Delta=0$, $V=0$ in $H_{1}$, and $V=0.1 \,$MHz in $H_{2}$ --- panel (b$_{1}$)],
    the alternative model of Sect.~\ref{sec:model2}
    [$\Delta=0.8 \,$MHz in $H_{1}$, $\Delta=0$ in $H_{2}$, and $V=0.1$ MHz --- panel (b$_{2}$)],
    and the original model of Sect.~\ref{sec:model}
    [$\Delta=0.8 \,$MHz and $V=0.1 \,$MHz --- panel (b$_{3}$)],
    plotted against the chain size $L$ and the perturbation $\epsilon$.
    The yellow, green, and blue regions respectively correspond to $\delta n_{c}>0$, $\delta n_{c}=0$,
    and $\delta n_{c}<0$.}
  \label{fig8}
\end{figure}

As a matter of fact, in Fig.~\ref{fig8}(a) we explicitly show what happens to the persistence of the DTC regime
for the model of Sect.~\ref{sec:model}, by focusing on several $\Delta$ values close to
the central peak at $\Delta=0.13 \,$MHz in Fig.~\ref{fig5}(c$_{1}$) with $\epsilon=0.4 \,$MHz.
It is clear that the exponential growth of $n_{c}$ does not continue indefinitely, and stable subharmonic oscillations
do not survive beyond a critical chain size $L_{c}$, which becomes
smaller for a larger deviation from the central peak, consistently with Fig.~\ref{fig5}(c$_{1}$).

To gain further insight on the range of parameters that are amenable to the observation of the DTC features,
we now assume that the increase of $n_c$ with $L$ being a fair numerical indicator of a putative DTC phase.
Bearing in mind this heuristic criterion, we have numerically calculated the discrepancy $\delta n_c \equiv n_c(L)-n_c(L-1)$
for a given set of parameters, and identified each point in the parameter space at a given size $L$
that results in $\delta n_c > 0$ as belonging to the DTC regime.
The resulting finite-system phase diagram is shown in the bottom panels of Fig.~\ref{fig8},
against the chain size $L$ and the perturbation $\epsilon$.
The three panels correspond to, respectively, the simplified model in Appendix~\ref{app:analytic}
($V=0$ in $H_{1}$, and $\Delta=0$) [panel (b$_{1}$)], the alternative model of Sect.~\ref{sec:model2} [panel (b$_{2}$)],
and the original model of Sect.~\ref{sec:model} [panel (b$_{3}$)].
It is easy to see that all the models exhibit a boundary
delimiting the parameter range which supports a persistent DTC regime (yellow region, characterized by $\delta n_c > 0$)
from the rest (green and blue regions, respectively characterized by $\delta n_c = 0$ and $\delta n_c < 0$),
being the DTC regime favored for $\left\vert \epsilon\right\vert $ small enough.
Including $V$ in $H_{1}$ and $\Delta$ in $H_{2}$ results in a moderate translation and/or a reduction
of the DTC regime along the $\epsilon$ axis, while introducing more frequent instances
of a non monotonic increase of $n_{c}$ with $L_{c}$ (blue regions), without however changing the overall picture.

Since in all our simulations the energy scale dictated by the Floquet frequency $2\pi/T$ of the external driving
is comparable to the intrinsic energy scales, and in particular to the scale $V$ of the many-body interactions,
we consider unlikely that the numerical evidence of DTC signatures extending to more than ten thousand cycles
[as in Fig.~\ref{fig4}(a$_3$) and Fig.~\ref{fig6}(a$_3$)] might be due to a prethermal time crystal,
even though we cannot completely rule out this possibility.
Were this to be the case, we might attribute the reduction of the favorable $\epsilon$ range
[yellow regions in Fig.~\ref{fig8}(b$_1$)-(b$_3$)] to a faster final thermalization in longer chains,
pre-empting the increase of $n_c$ with $L$.

%%%%%%%%%%%%%%%%%%%%%%%%%%%%%%%%%%%%%%%%%%%%%%%%%%%%%%%%%%%%%%%%%%%%%%%%%%%%%%%%%%%%%%%
%%%%%%%%%%%%%%%%%%%%%%%%%%%%%%%%%%%%%%%%%%%%%%%%%%%%%%%%%%%%%%%%%%%%%%%%%%%%%%%%%%%%%%%

\section{Dissipation effects}
\label{sec:dissipation}

While the models we are interested in are amenable to a direct implementation in the lab, and the range of chain sizes we
numerically simulate ($L \sim 10$) are clearly of experimental interest, so far we have disregarded the fact that
the Rydberg state has a long, but finite, lifetime that will unavoidably introduce a source of dissipation in the system.
In order to assess how the DTC features discussed above are possibly affected by the decay rate $\Gamma$
of the Rydberg state $|r\rangle_j$ of each atom, we resort to a Markovian master equation treatment~\cite{Gardiner},
describing the time evolution of the density matrix $\rho$ of the $L$-atom system in the Lindblad form:
\begin{equation}
  \partial_{t}\rho = -\frac{i}{\hbar}[H(t),\rho] + \mathcal{D}(\rho) .
  \label{master}
\end{equation}
The incoherent dissipation is described by the term
\begin{equation}
  \mathcal{D}(\rho) = \Gamma \sum_{j=1}^L \left[ \sigma_{j}^{-} \rho \sigma_{j}^{+}
    - \frac{1}{2} \big( \rho \sigma_{j}^{+} \sigma_{j}^{-} + \sigma_{j}^{+} \sigma_{j}^{-} \rho \big) \right]
\end{equation}
associated with the jump operators $L_j = \sqrt{\Gamma} \sigma^-_j$,
describing dissipation processes.
Once $\rho(t)$ is attained for our Rydberg chain (see Appendix~\ref{app:numerics} on details), it is then straightforward to
calculate the population imbalance as $P(t) = \tfrac{1}{L} \sum_{j=1}^L {\rm Tr} \big[ \rho(nT) N_{j} \big]$.

%%%%%%%%%%%%%%%%%%%%%%%%%%%%%%%%%%%%%%%%%%%%%%%%%%%%%%%%%%%%%%%%%%%%%%%%%%%%%%%%%%%%%%%
\begin{figure}[!t]
  \includegraphics[width=8.0cm]{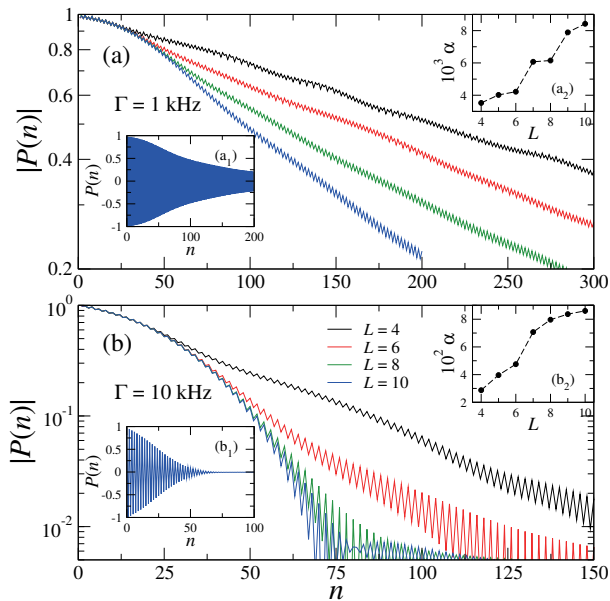}\caption{(Color online) The modulus of the average population imbalance
    $\left\vert P(n)\right\vert $ against the Floquet cycle number $n$ in semilog scale,
    with decay rates $\Gamma=1.0 \,$kHz (a) and $\Gamma=10 \,$kHz (b).
    Black, red, green, and blue lines correspond to $L=4$, $L=6$, $L=8$, and $L=10$, respectively.
    The other parameters are chosen as $\epsilon = -0.1 \,$MHz, $V = 0.1 \,$MHz, $\Delta = 0$,
    $T_{1} = 1.0 \, \mu$s, and $T_{2}=15 \, \mu$s.
    The two left insets (a$_1$) and (b$_1$) show the same time traces of the corresponding main frames for $L=10$,
    without taking the modulus of $P(n)$.
    The two right insets (a$_2$) and (b$_2$) show the decay rates of the population imbalance as a function of $L$,
    obtained as results of an exponential fit $|P(n)| \sim e^{- \alpha n}$ of the numerical data.}
  \label{fig9}
\end{figure}
%%%%%%%%%%%%%%%%%%%%%%%%%%%%%%%%%%%%%%%%%%%%%%%%%%%%%%%%%%%%%%%%%%%%%%%%%%%%%%%%%%%%%%%

The outcomes of our simulations for $\Delta = 0$ are reported in Fig.~\ref{fig9}, where we show
a few time traces of the polarization amplitude $|P(n)|$ for different values of the chain size $L$,
in the presence of a nonzero $\Gamma$, by directly solving Eq.~\eqref{master}.
It is easy to see that, though $|P(n)|$ decays faster or slower for a larger or smaller $\Gamma$
[compare the decay rates of panel (a), for $\Gamma = 1\,$kHz, with those of panel (b), for $\Gamma = 10\,$kHz],
the fixed frequency oscillation can be observed (left insets) as a key DTC feature
for tens or hundreds of Floquet cycles on the millisecond scale. It is also
clear that $|P(n)|$ typically suffers a faster decay for a larger $L$, until it
reaches a saturation value [cf., the green and blue lines in panel (b)].
For illustrative purposes, we have tried and fit our numerical data for the curves of $|P(n)|$ decaying with $n$ using a single exponential:
$|P(n)| \sim e^{- \alpha n}$, thus obtaining the decay rates $\alpha$ plotted in the two right insets.

Similar results to those just discussed on the effects of the finite lifetime of the Rydberg level are also obtained for non vanishing values of $\Delta$ and for both models above (Sect.s~\ref{sec:model} and \ref{sec:model2}).

\section{Conclusions}
\label{sec:concl}

In summary, we have numerically investigated two realistic models exhibiting DTC features based on a ring of cold
Rydberg atoms interacting via van der Waals potentials.
Our results show that, by applying suitable sequences of driving fields with a Floquet periodicity $T$, it is viable to fulfill all three conditions required for DTCs. The population difference oscillates with a period $2T$ ({\it discrete time-translational symmetry breaking}), such a period is robust against parameter variations ({\it rigidity}), the oscillations remain in phase over an increasingly long time when the chain length increases ({\it persistence}). Employing an exact diagonalization approach, we can deal with chain sizes of up to 14 atoms, which may definitely be of experimental interest. We have characterized the parameter ranges amenable to the observation of the DTC regime for two distinct set-ups. While these share many common features, such as the symmetry for a simultaneous sign change of both the driving detuning $\Delta$ and the
dipole-dipole interaction $V$, they are differently sensitive to the value of $\Delta$. Finally, we have considered the effects of a finite lifetime of the Rydberg level which may limit the persistence of the DTC features for realistic values of $\Gamma$. In this regards, however, we stress that all other parameters considered here widely tunable: the Rabi frequency $\Omega$ and detuning $\Delta$ being directly controlled via the intensity and frequency of the driving laser fields, while the interaction parameter $V$ by the distance between the adjacent potential wells in the chain. Thus, the models we consider here are amenable to experimental verification even in the case in which extrinsic effects, such as noise associated to the laser stability or other dephasing effects, might
effectively lead to  somewhat larger values of $\Gamma$.

\acknowledgments

The work is supported by the National Natural Science Foundation of China
(No.~10534002 and No.~11674049) as well as the Cooperative Program by the Italian
Ministry of Foreign Affairs and International Cooperation (No.~PGR00960) and
the National Natural Science Foundation of China (No.~11861131001).

%\newpage

\appendix

\section{Numerical approach}
\label{app:numerics}

In order to study the emergence of the DTC features in our Rydberg-chain system,
we resorted to an exact-diagonalization (ED) approach.
Below we comment on the applicability of other routinely employed numerical methods,
in order to probe the Floquet dynamics of quantum many-body systems.

Let us start from purely Hamiltonian situations (i.e., we neglect dissipation for a moment).
For the sake of clarity, here we explicitly refer to the model of Sect.~\ref{sec:model};
however, from a computational point of view, the employed procedure
and the complexity of calculations are totally equivalent in the other considered models.
Our simulations make use of the full $2^L$-dimensional Hilbert space of the system, $L$ being the number
of Rydberg atoms in the simulated chain.
We start from a given initial state $|\Psi(0)\rangle$ in such space, written in the basis
$\{ |g\rangle_j, |r\rangle_j \}_{j=1, \ldots L}$, and repeatedly apply the Floquet
time-evolution operator
\begin{equation}
  U_F(1) = U_2 U_1 = e^{-i H_2 T_2 / \hbar} e^{-i H_1 T_1/\hbar}  \nonumber
\end{equation}
as in Eq.~\eqref{evoop}.
While $U_2$ is already diagonal in the above computational basis, and thus can be trivially applied to
a generic input state with $O(2^L)$ operations, the evaluation of the nondiagonal operator $U_1$ generally
requires the full diagonalization of the matrix $H_1$ in Eq.~\eqref{H1}.
Unfortunately, besides the exponential growth of the system's Hilbert space, the need to calculate
and manipulate all the eigenvalues and eigenvectors of $H_1$ requires to act on $O(2^{2L})$ complex elements,
thus severely limiting present-day numerical capabilities up to $L \sim 15$ atoms.

In principle, other less computationally demanding, yet approximate, methods could be used in order to track
the quantum dynamics of the system. For example, one could either resort to a numerical integration
of the Schr\"odinger equation governing the time evolution of the system (either via a standard Runge-Kutta approach
or with a Suzuki-Trotter decomposition of the unitary evolution operator), or to techniques based
on the density-matrix renormalization group (DMRG).
The first type of approaches still requires to manipulate and keep track of the full $2^L$-dimensional Hilbert space,
although the full Hamiltonian spectrum is not explicitly used (taking advantage of the sparseness
of the various Hamiltonian matrices, one would be able to deal with sizes of $L \sim 30$ atoms).
However we checked that the need to reach very long times, up to several thousands of Floquet time intervals
(see, e.g., Figs.~\ref{fig4} and~\ref{fig6} where $nT \gtrsim 0.1 \,$s),
in order to monitor the persistence properties of the DTC phase,
limits the applicability of approximate methods, due to accumulation of numerical errors.
Likewise, the entanglement growth along the Floquet time evolution prevents DMRG-based algorithms
to achieve such regimes, although DTC signatures for a relatively small number of Floquet periods
and much larger systems have been reported in a similar context~\cite{Huang}.

Finally we briefly discuss the simulation procedure adopted for a non-unitary dynamics in the presence of losses,
as described by the master equation~\eqref{master}.
Even in this case we employed an ED method, allowing to reach arbitrarily large Floquet cycles
without error accumulation. To this purpose, we first vectorize the system's density matrix
\begin{equation}
  \rho = \sum_{{\bm i}, {\bm j}} \rho_{{\bm i}, {\bm j}} |{\bm i}\rangle \langle {\bm j}| \; \longrightarrow \;
  |\rho\rangle\!\rangle \equiv \sum_{{\bm i}, {\bm j}} \rho_{{\bm i}, {\bm j}} |{\bm i}\rangle \otimes |{\bm j}\rangle ,
  \label{eq:rhomap}
\end{equation}
with ${\bm i} \equiv \{ i_1, \ldots, i_L \}$ (and $i_\ell$ denoting the state of the $\ell$-th Rydberg atom),
and then formally write the master equation as a linear differential equation on a vectorized state of $2^{2L}$
components (analogously to the Schr\"odinger equation, which acts on a pure state of $2^L$ components):
\begin{equation}
  \partial_t |\rho\rangle\!\rangle = \mathcal{L}(t) |\rho\rangle\!\rangle.
\end{equation}
Here $\mathcal{L}(t)$ denotes the Liouvillian superoperator, of size $2^{2L} \times 2^{2L}$,
applied to the vectorized state $|\rho\rangle\!\rangle$,
and obtained from the master equation through the same mapping as in Eq.~\eqref{eq:rhomap}.
Similarly as for the Hamiltonian case, the ED approach deals with the full Liouvillian spectrum,
thus acting on $O(2^{4L})$ complex elements;
for the times we were interested in, we were able to reach systems with up to $L=10$ sites.

\section{Few-bodies and -cycles dynamics}
\label{app:analytic}

In this appendix, we show explicit analytical results on $P(n)$ in the few-cycles and few-bodies regime,
with the purpose to gain a qualitative understanding of the dependence on parameters of the DTC features shown in
Figs.~\ref{fig4},~\ref{fig5},~\ref{fig6}, and~\ref{fig7}.
Let us introduce a simplified model, for which it is possible to derive an analytic solution in a relatively compact form,
and set $V=0$ during $T_{1}$ such that only Rabi flipping and detuning terms
are admitted in the first stage of the Floquet cycle. The corresponding Hamiltonian is
\begin{equation}
  \tilde H_{1} = \hbar \sum \limits_{j=1}^{L}
  \left[ \Omega\left( \sigma_{j}^{+} + \sigma_{j}^{-} \right) + \Delta N_{j}^{r} \right] .
\end{equation}
On the other hand, both detuning and interaction terms are present in the second stage,
which is thus described by the Hamiltonian $H_2$ in Eq.~\eqref{H2}.

In the case of two atoms, the stroboscopic evolution operator $\widetilde U_F = U_2 \tilde U_1$ is obtained by composing
the following explicit matrix representation
\begin{equation}
  \tilde U_{1} \! = \! \left( \! \!
  \begin{array}
    [c]{cccc}%
    X_{+}^{2}e^{i2\theta_{1}} & iX_{+}Ye^{i2\theta_{1}} & iX_{+}Ye^{i2\theta_{1}} & -Y^{2}e^{i2\theta_{1}}\\
    iX_{+}Y & X^{2} & -Y^{2} & iX_{-}Y\\
    iX_{+}Y & -Y^{2} & X^{2} & iX_{-}Y\\
    -Y^{2}e^{-i2\theta_{1}} & iX_{-}Ye^{-i2\theta_{1}} & iX_{-}Ye^{-i2\theta_{1}} & X_{-}^{2}e^{-i2\theta_{1}}
  \end{array}
  \! \! \right)
  \label{matrU1}
\end{equation}
and
\begin{equation}
  U_{2} = \left(
  \begin{array}
    [c]{cccc} 1 & 0 & 0 & 0\\
    0 & e^{-i\theta_{2}} & 0 & 0\\
    0 & 0 & e^{-i\theta_{2}} & 0\\
    0 & 0 & 0 & e^{-i(2\theta_{2}+\theta_{3})}
  \end{array}
  \right) ,
  \label{matrU2}
\end{equation}
where
\begin{align}
  &X_{\pm} =  \cos(\Omega_{e}T_{1})\pm i(\Delta/\Omega_{e})\sin(\Omega_{e}T_{1}), \nonumber \\
  &Y =  (\Omega/\Omega_{e})\sin(\Omega_{e}T_{1}), \nonumber \\
  &X = \sqrt{X_{+}X_{-}}, \nonumber \\
  &\theta_{1} = \Delta T_1/2, \quad \theta_{2} = \Delta T_{2}, \; \mbox{ and } \; \theta_{3} = VT_{2}. \nonumber
\end{align}
The operator $\widetilde U_F$ acts on the $4\times1$ column vector obtained by arranging the two-atom
state basis $\left\vert gg\right\rangle $, $\left\vert gr\right\rangle $,
$\left\vert rg\right\rangle $, and $\left\vert rr\right\rangle $ in this order. Thus,
\begin{equation}
  \left\vert \Psi(n)\right\rangle = \widetilde U_{F}(n)\left\vert \Psi(0)\right\rangle
  = \big[ U_{2} \tilde U_{1} \big]^{n}\left\vert \Psi(0)\right\rangle.
\end{equation}
Assuming that the atomic evolution starts from the ground state, i.e.,
that its components on the basis used in the above matrix representations~\eqref{matrU1} and~\eqref{matrU2}
are $\Psi_{1}(0)=1$ and $\Psi_{2}(0)=\Psi_{3}(0)=\Psi_{4}(0)=0$, we then have
$P(0)=-1$ with $P(n)=\Psi_{4}^{\ast}(n)\Psi_{4}(n)-\Psi_{1}^{\ast}(n)\Psi_{1}(n)$.
After two Floquet cycles, we obtain
\begin{widetext}
  \begin{equation}
    P(2) = -X^{8}+2X^{4}Y^{4}-Y^{8}
    + \Big[ 2X_{+}^{2}Y^{4}(Y^{2}+X^{2})e^{i(\varphi_{1}+\theta_{3})}
      + 2X_{+}^{2} X^{2}Y^{2}(Y^{2}+X^{2})e^{i\varphi_{1}} + {\rm c.c.} \Big] ,
  \end{equation}
  where $\varphi_{1}=2\theta_{1}+\theta_{2}$. After the third cycle, we have
  \begin{align}
    P(3) = & -X^{12}-5X^{8}Y^{4}-16X^{6}Y^{6}-11X^{4}Y^{8}+ Y^{12} \nonumber\\
    & + \Big[ 2X_{+}^{6}Y^{4}\left( X^{2}+Y^{2}\right) e^{i(3\varphi_{1}+2\theta_{3})}
      + 2X_{+}^{4}Y^{4}\left( X^{4}-Y^{4} \right) e^{i(2\varphi_{1}+2\theta_{3})}
      - 2X_{+}^{6}Y^{4}\left( X^{2}+Y^{2} \right) e^{i(3\varphi_{1}+\theta_{3})} \nonumber\\
      & \; \quad + 4X_{+}^{4}Y^{4} \left( X^{4}+2X^{2}Y^{2}+Y^{4} \right)
      e^{i(2\varphi_{1}+\theta_{3})} + 2X_{+}^{2}Y^{4}\left( 4X^{6}-6X^{4}Y^{2}+9X^{2}Y^{4}-Y^{6} \right)
      e^{i(\varphi_{1}+\theta_{3})} \nonumber\\
      & \; \quad + 2X_{+}^{4}X^{2}Y^{2} \left( X^{4}-Y^{4} \right) e^{2i\varphi_{1}}
      + 2X_{+}^{2}Y^{2} \left( 2X^{8}+8X^{6}Y^{2}-10X^{4}Y^{4}+7X^{2}Y^{6}-Y^{10} \right) e^{i\varphi_{1}} \nonumber\\
      & \; \quad + 2X^{2}Y^{4} \left( X^{6}+X^{4}Y^{2}-X^{2}Y^{4}-Y^{6} \right) e^{i\theta_{3}} + {\rm c.c.} \Big],
    \label{A7}
  \end{align}
  which is sensitive to $\varphi_{1}$ (that is, to $\Delta$) and $\theta_{3}$ (that is, to $V$)
  in a more complicated way than $P(2)$.

  In an analogous way, in the case of three atoms for which $\tilde U_1$ and $U_2$ become $8\times 8$ matrices, we obtain
  \begin{align}
    P(2) = & -\left(  X^{2}-Y^{2}\right)  ^{2}\left( Y^{2}+X^{2}\right)^{4} \nonumber \\
    & + \Big[ 2X_{+}^{4}Y^{6}\left( 2X^{2}-Y^{2}\right) e^{i(2\varphi_{1}+3\theta_{3})}
      + 2X_{+}^{2}Y^{6}\left( Y^{2}+X^{2}\right)^{2} e^{i(\varphi_{1}+2\theta_{3})}\nonumber\\
    & \; \quad + 4X_{+}^{2}X^{2}Y^{4}\left( Y^{2}+X^{2}\right) e^{i(\varphi_{1}+\theta_{3})}
      + 2X_{+}^{2}X^{4}Y^{2} \left( Y^{2}+X^{2}\right)^{2}e^{i\varphi_{1}} + {\rm c.c.} \Big] .
    \label{A13}
  \end{align}
\end{widetext}

If the signs of both $\Delta$ and $V$ are
changed, $X$ and $Y$ are invariant while $X_+$ and $X_-$ exchange role, while the phases $\varphi_{1}$ and $\theta_{3}$
and their linear combinations change sign. As in all the above expressions for $P(n)$
all complex terms are summed with their complex conjugate ones, $P(n)$ remains the same.
In other terms, the contributions to $P(n)$ contained in the square brackets of the above equations reduce to a sum of terms of the form
\begin{equation}
  f_{1}(\Delta^{2j})\cos[g_{1}(\Delta,V)]+f_{2}(\Delta^{2j-1})\sin[g_{2}(\Delta,V)] \nonumber
\end{equation}
with $f_{1,2}$ and $g_{1,2}$ denoting certain linear functions, and thus $P(n)$ is invariant
when the signs of $\Delta$ and $V$ are simultaneously changed
[as in panels (c$_1$) and (c$_2$) of Figs.~\ref{fig5} and \ref{fig7}].

We also note that the interference phase terms involve increasingly varied combinations of
$\varphi_{1}=\Delta(T_{1}+T_{2})$ and $\theta
_{3}=VT_{2}$  as the Floquet cycle $n$ and/or the chain size $L$
increase. The number of such linear combinations grows exponentially as $2^{L-1}4^{n-2}$
for $L\geq2$ Rydberg atoms and $n\geq2$ Floquet cycles, while the oscillations of $X_\pm$ and $Y$ only depend on
the phase $\alpha=\Omega_e T_1$. It is to be noticed that in the model of Sect.~\ref{sec:model} the value of
$\varphi_1$  is much more sensitive to $\Delta$ than is $\alpha$ as $T_2\gg T_1$, while this is not the case for the model of Sect.~\ref{sec:model2}.

\end{document}